# Energy-Efficient Distributed Machine Learning in Cloud Fog Networks

Mohammed M. Alenazi
School of Electronic and Electrical
Engieering, University of Leeds, Leeds
elmmal@leeds.ac.uk

Barzan A. Yosuf
School of Electronic and Electrical
Engieering, University of Leeds, Leeds
b.a.yosuf@leeds.ac.uk

Sanaa H. Mohamed
School of Electronic and Electrical
Engieering,University of Leeds, Leeds
s.h.h.mohamed@leeds.ac.uk

Taisir E.H. El-Gorashi
School of Electronic and Electrical
Engieering,University of Leeds, Leeds
t.e.h.elgorashi@leeds.ac.uk

Jaafar M. H. Elmirghani
School of Electronic and Electrical
Engieering, University of Leeds, Leeds
j.m.h.elmirghani@leeds.ac.uk

*Abstract*—Massive amounts of data are expected to be generated by the billions of objects that form the Internet of Things (IoT). A variety of automated services such as monitoring will largely depend on the use of different Machine Learning (ML) algorithms. Traditionally, ML models are processed by centralized cloud data centers, where IoT readings are offloaded to the cloud via multiple networking hops in the access, metro, and core layers. This approach will inevitably lead to excessive networking power consumptions as well as Quality-of-Service (QoS) degradation such as increased latency. Instead, in this paper, we propose a distributed ML approach where the processing can take place in intermediary devices such as IoT nodes and fog servers in addition to the cloud. We abstract the ML models into Virtual Service Requests (VSRs) to represent multiple interconnected layers of a Deep Neural Network (DNN). Using Mixed Integer Linear Programming (MILP), we design an optimization model that allocates the layers of a DNN in a Cloud/Fog Network (CFN) in an energy efficient way. We evaluate the impact of DNN input distribution on the performance of the CFN and compare the energy efficiency of this approach to the baseline where all layers of DNNs are processed in the centralized Cloud Data Center (CDC).

## I. INTRODUCTION

Machine learning (ML) is increasingly used in many fields such as medical applications, smart cities, and autonomous cars where the goal is to efficiently and accurately predict the output or best response to new input data [1]. Nowadays, massive amounts of data can be produced by distributed Internet-of-Things (IoT) devices which reached more than 50 billion [2]. Using the abundant IoT data, intelligent services can be provided at edge networks such as distributed detection, monitoring, and classification [3] where Deep Neural Networks (DNNs) are one of the most widely used ML tools for these applications. Traditionally, due to computational complexity, ML algorithms used to be processed in centralized cloud data centers (CDCs). While it is evident that the usage of centralized data centres for ML has provided accuracy and high performance, nevertheless it is achieved at the cost of high energy consumption [3], [4]. Transferring input data to CDCs imposes networking overheads in terms of power consumption and delay and also raise privacy concerns as the data could be accessed for unauthorized purposes [5].

To address the aforementioned delay and power consumption challenges, researchers have proposed various distributed processing tools such as fog computing where a large number of devices such as IoTs and different "fog nodes" with excess processing, memory and networking resource can process the data partially closer to the data sources [6], [7], [8]. With a focus on the energy efficiency of ML, we proposed a decentralized solution where ML algorithms for service requests by a single IoT are embedded in different layers on a cloud/fog architecture [9]. We address in this paper the optimization of virtualizing and embedding the resources required for DNNs on the cloud/fog architecture where the data sources are distributed IoTs. This work progresses our previous work on the energy efficiency of networking and computing for distributed processing for IoT [10]-[13], virtual machine placement [14], [15], virtual network embedding [16], content distribution [17]-[23], big data applications [24]-[29], and also designing optical core networks and future data center networks [30]-[39].

As the applications using DNNs increase in their computational complexity, their associated energy consumption becomes challenging. In the case of edge/ fog computing, such challenges heighten because the edge devices are resource constrained as they operate on a limited energy budget [4]. In the literature, the energy efficiency of ML algorithms, specifically deep learning models was tackled on a number of levels, (i) improving the algorithms so that the number of multiplication-and-accumulations (MACs) is minimized in the code [1], [5], (ii) performing specialized optimization at the hardware level e.g., using high-end Graphical Processing Units (GPUs), and Application-Specific Integrated Circuits (ASICs) [6], (iii) distributing the hidden layers across heterogeneous processing resources offered by Cloud/Fog networks [7], [8], [9]. In this paper we take the approach in (iii) by designing a cross-layer optimization framework that efficiently allocates virtualized functions (i.e., layers) of generic DNN models across heterogenous layers of processing offered by a cloud/fog network architecture.

The remainder of this paper is organized as follows: Section II describes the cloud/fog system model and the optimization model for placing DNNs in the cloud/fog architecture. Section III provides the results and discussions while Section IV provides the conclusions and future work.

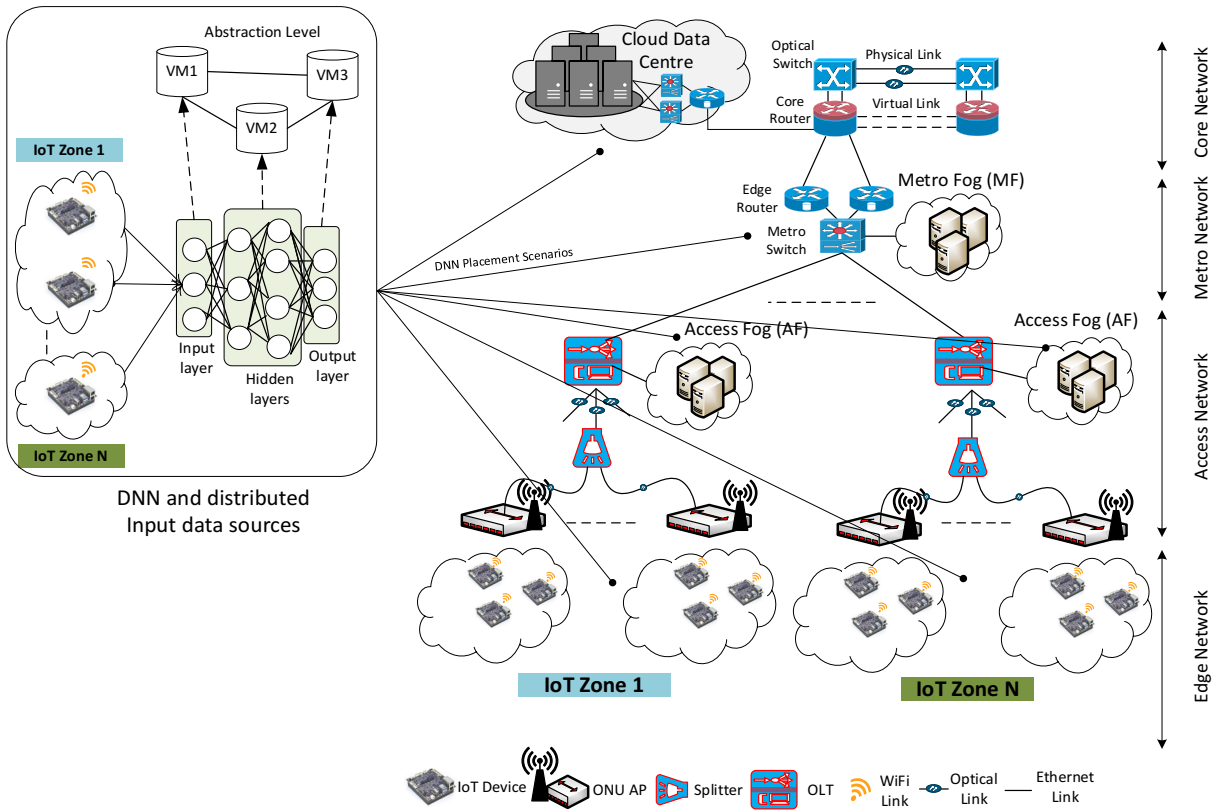

Figure 1 Proposed cloud/fog network with distributed IoT devices requesting DNNs.

## II. SYSTEM MODEL

In this section, we introduce the cloud/fog architecture evaluated and explain the role of the virtualization approach in embedding the VSRs, which represent different DNN models. We conclude this section with the problem formulation, where we present the notations used, key mathematical equations in terms of the objective function and constraints.

### A. Cloud/fog network architecture and potential processing locations

Figure 1 shows the Cloud/Fog Network (CFN) we considered. It mainly comprises of four networking layers which are edge, access, metro, and core networks. In the edge network, we consider distributed IoT devices in different zones where they could collect different types of data. For the access layer, we consider a Passive Optical Network (PON) in each IoT zone. Each PON contains several Optical Line Terminals (ONU) that connect with the IoT devices through Wi-Fi and aggregate the traffic from IoTs via fiber links and a splitter into a future-proof Optical Line Terminal (OLT) [40]. An Access Fog (AF) containing several servers is connected to each OLT (i.e., in each IoT zone). The OLT connects to the metro network through a metro switch which connects to the core network via a metro edge router. In the metro network, a Metro Fog (MF) is considered containing a set of servers. The core network we consider is an IP over WDM network [17] which has two layers, an IP layer and an optical layer. The IP layer in each core node connects to the metro network and aggregates its uplink and downlink traffic and the optical layer performs electrical to optical / optical to electrical conversion and physically connects to other core nodes via optical fiber links. We consider a central Cloud Data Center (CDC) connected to a core node that is one hop from the core node aggregating from the metro and access networks that link the considered IoTs.

In this network, ML inference for requests is issued by distributed IoTs. These requests can be processed in the IoT end-devices, Access Fog (AF), Metro Fog (MF), and the centralized Cloud Datacenter (CDC). This can be referred to as IoT-Cloud continuum [41]. Given the benefits of hardware and network virtualization, which include resources abstraction, consolidation, and isolation between different users, we consider a fully virtualized CFN architecture. A single DNN algorithm, which we call a Virtual Service Request (VSR) can then be modeled as multiple Virtual Machines (VMs) that can be placed at the aforementioned processing locations, interconnected through a virtual network topology as illustrated in Figure 1. Different DNN algorithms can be represented by random virtual topologies to be embedded onto the CFN architecture [9]. For a certain VSRs processing requirement, a trade-off between the processing efficiency and networking power consumption for the inter VM traffic and input data transfer arises when selecting the optimal VM location in terms of the energy efficiency of ML. In the following subsection, we formulate a Mixed Integer Linear Programming Model (MILP) that optimizes the placement of the VMs of each VSR and the embedding of their virtual networks in the CFN architecture.

### B. Problem Formulation using MILP

Benefiting from our track record in MILP optimization and particularly in network virtualization and service embedding in [13], [14], respectively, we developed a model to optimize the placement of virtual DNN functions (or VSRs) in the CFN. VSRs comprise of multiple VMs, each VM represents a layer of a DNN model that has a demand for processing (in Floating Point Operation per Second (FLOPS)) and networking (in Mbps). Consequently, a VSR is embedded optimally on the CFN model while respecting the capacity constraints of processing and networking devices. The physical network shown in Figure 1 is modelled as an undirected graph $G = (N, L)$, where $N$ represents the set of all nodes and $L$ the set of links connecting those nodes in the topology. The VSR $s$ is represented by the directed graph $G^r = (R^r, L^r)$, where $R^r$ is the set of VMs representing virtualized DNN layers and $L^r$ is the set of virtual links connecting those VMs. In [9], we exemplify how demands in a VSR are mapped onto the physical resources in the CFN architecture and show clearly one of the key variables used to establish the virtual links to achieve the inter-VM communication. Before introducing the optimization model, we define the sets, parameters and variables used:

Sets:
- $\mathbb{DC}$ — Set of centralized data centres (CDC).
- $\mathbb{C}$ — Set of IP/WDM nodes in the core network.
- $\mathbb{MF}$ — Set of metro fog (MF) nodes.
- $\mathbb{AF}$ — Set of access fog (AF) nodes.
- $\mathbb{O}$ — Set of Wi-Fi enabled ONU access points.
- $\mathbb{I}$ — Set of IoT end-devices.
- $\mathbb{P}$ — Set of processing nodes that can host virtual service requests (VSRs), where $\mathbb{P} = \mathbb{DC} \cup \mathbb{MF} \cup \mathbb{AF} \cup \mathbb{I}$.
- $\mathbb{R}$ — Set of virtual service requests (VSRs).
- $\mathbb{VM}_r$ — Set of virtual machines (VMs) in VSR $r \in \mathbb{R}$.
- $\mathbb{IP}$ — Set of IoT end-devices that act as input, where $\mathbb{IP} \subset \mathbb{I}$.
- $\mathbb{N}$ — Set of all nodes in the proposed CFN architecture.
- $\mathbb{N}_m$ — Set of neighbor nodes of node $m \in \mathbb{N}$.

Parameters:
- $s$ and $d$ — Index of the source and destination nodes of a VSR topology.
- $b$ and $e$ — Index of the source and destination after VMs are processed by processing nodes $b, e \in P, b \neq e$.
- $m$ and $n$ — Index the physical links in the CFN topology.
- $F^{r,s}$ — Processing demand by node $s$ in VSR $r$, in FLOPS.
- $H^{r,s,d}$ — Bitrate demand by VSR $r$ on the virtual link $(s, d) \in \mathbb{VM}_r$.
- $P_s^r$ — $P_s^r = 1$, if in VSR $r \in \mathbb{R}$, virtual machine $s \in VMr$ is input, otherwise $P_s^r = 0$.
- $\Pi_n^{(net)}$ — Maximum power consumption of network node $n \in \mathbb{N}$.
- $\pi_n^{(net)}$ — Idle power consumption of network node $n \in \mathbb{N}$.
- $\epsilon_n$ — Power per Gb/s of network node $n \in \mathbb{N}$.
- $\Pi_p^{(LAN)}$ — Maximum power consumption of LAN network inside processing node $p \in \mathbb{P}$.
- $\pi_p^{(LAN)}$ — Idle power consumption of networking equipment inside processing node $p \in \mathbb{P}$.
- $EL_p$ — Energy per bit of network node $n \in \mathbb{N}$, in W/(Gb/s).
- $\Pi_p^{(pr)}$ — Maximum power consumption of a single processing server at node $n \in \mathbb{P}$.
- $\pi_p^{(pr)}$ — Idle power consumption of a single processing server at node $p \in \mathbb{P}$.
- $NS_p$ — Maximum number of servers deployed at processing node $p \in \mathbb{P}$.
- $E_p$ — Energy per GFLOPS of processing node $p \in \mathbb{P}$.
- $\delta$ — Proportion of idle power consumed on high-capacity networking equipment.
- $PUE_n^{(net)}$ — Power Usage Effectiveness (PUE) of node $n \in N$ for networking.
- $PUE_p^{(net)}$ — Power Usage Effectiveness (PUE) of node $p \in P$ for processing.

Variables:
- $\lambda^{b,e}$ — Traffic demand between processing node pair $(b, e) \in \mathbb{P}$ aggregated after all VSRs are embedded.
- $\lambda_{m,n}^{b,e}$ — Traffic demand between processing node pair $(b, e) \in \mathbb{P}$ aggregated after all VSRs are embedded, traversing physical link $(m, n), m \in \mathbb{N}$ and $n \in \mathbb{N}_m$.
- $\lambda_n$ — Amount of traffic aggregated by network node $n \in \mathbb{N}$, where
$\lambda_n = \sum_{b \in \mathbb{P}} \sum_{e \in \mathbb{P}:b \neq e} \sum_{m \in \mathbb{N}} \sum_{n \in \mathbb{N}_m} \lambda_{m,n}^{b,e} + \sum_{b \in \mathbb{P}} \sum_{e \in \mathbb{P}:b \neq e} \sum_{m \in \mathbb{N}:m \neq e} \sum_{n \in \mathbb{N}_m} \lambda_{n,m}^{b,e}$.
- $\beta_n$ — $\beta_n = 1$, if network node $n \in \mathbb{N}$ is activated, otherwise $\beta_n = 0$.
- $\theta_p$ — Amount of traffic aggregated by processing node $p \in \mathbb{P}$.
- $\Omega_p$ — Amount of workload in FLOPS, allocated to processing node $p \in \mathbb{P}$.
- $N_p$ — Number of activated processing servers at processing node $p \in \mathbb{P}$.
- $\Phi_p$ — $\Phi_p = 1$, if processing node $p \in \mathbb{P}$ is activated, otherwise $\Phi_p = 0$.
- $\delta_b^{r,s}$ — $\delta_b^{r,s} = 1$, if virtual machine $s \in VM_r$ is embedded for processing at node $b \in P$, otherwise $\delta_b^{r,s} = 1$.

Therefore, the total power consumption is made up of two parts: 1) total network power consumption, 2) total processing power consumption. It is important to note that processing power consumption includes the power consumed by the servers as well as the switches routers within these nodes to provide the LAN.

- Network power consumption, which is given by:

$$PUE_n^{(net)} \cdot \left[ \sum_{n \in \mathbb{N}} \epsilon_n \cdot \lambda_n + \sum_{n \in \mathbb{N}} \beta_n \cdot \pi_n^{(net)} \right] \quad (1)$$

The power consumption of the networking equipment comprises of power consumption of routers and switches of all the nodes in the CFN topology depicted in Figure 1 multiplied by the corresponding PUE at each node.

- Processing power consumption, which is given by:

$$PUE_p^{(pr)} \cdot \left[ \sum_{p \in \mathbb{P}} E_p \cdot \Omega_p + \sum_{p \in \mathbb{P}} N_p \cdot \pi_p^{(pr)} + \sum_{p \in \mathbb{P}} EL_p \cdot \theta_p + \sum_{p \in \mathbb{P}} \Phi_p \cdot \pi_p^{(LAN)} \right] \quad (2)$$

The first term of (2) is the proportional (or dynamic) power consumption of the servers whilst the second term calculates the idle (or static) power consumption of these servers. The third and fourth terms are the proportional and idle powers consumed by the internal LAN of the processing nodes, respectively.

Thus, the objective of the MILP is to minimize the total power consumption of the whole network as follows:

**Minimize:**

$$PUE_n^{(net)} \cdot \left[\sum_{n \in \mathbb{N}} \epsilon_n \cdot \lambda_n + \sum_{n \in \mathbb{N}} \beta_n \cdot \pi_n^{(net)}\right] +$$

$$PUE_p^{(pr)} \cdot \left[\sum_{p \in \mathbb{P}} E_p \cdot \Omega_p + \sum_{p \in \mathbb{P}} N_p \cdot \pi_p^{(pr)} + \sum_{p \in \mathbb{P}} EL_p \cdot \theta_p + \sum_{p \in \mathbb{P}} \Phi_p \cdot \pi_p^{(LAN)}\right]$$

**Subject to:**

$$\sum_{b \in \mathbb{P}} \delta_b^{r,s} = 1 \qquad \forall r \in \mathbb{R}, s \in \mathbb{VM}_r: P_s^r \neq 1 \qquad (3)$$

Constraint (3) ensures that all VMs within VSRs are processed, except for input VMs, as these must be processed by IoT source nodes.

$$\sum_{b \in \mathbb{IP}} \sum_{\substack{s \in \mathbb{VM}_r: \\ P_s^r = 1}} \delta_b^{r,s} = 1 \qquad \forall r \in \mathbb{R} \qquad (4)$$

Constraint (4) ensures that all input VMs are processed at IoT end-devices that act as source/input nodes.

$$\sum_{n \in \mathbb{N}_m} \lambda_{m,n}^{b,e} - \sum_{n \in \mathbb{N}_m} \lambda_{n,m}^{b,e} = \begin{cases} \lambda^{b,e} & m = s \\ -\lambda^{b,e} & m = d \\ 0 & otherwise \end{cases} \qquad (5)$$

$\forall b, e \in \mathbb{P}, d \in \mathbb{P}, m \in \mathbb{N}: b \neq e.$

Constraint (5) is the flow conservation constraint, which preserves the flow of traffic in the network.

Due to space limitations in this paper, only the key constraints are listed. The remainder of the constraints deal with binary indicators, capacity constraints on processing and networking devices and constraints that achieve the virtual embedding of VSRs.

Table 1: Servers input data for the MILP model.

| Devices | Max(W) | Idle(W) | GFLOPS | Efficiency (W/GFLOPS) |
|---|---|---|---|---|
| IoT (Rpi 4 B 4GB) | 7.3 [47] | 2.56 [47] | 13.5 [47] | 0.35 |
| AF Server (Intel i5-3427U) | 32.6 [47] | 10 [47] | 47.7 [47] | 0.47 |
| MF Server (Intel i5-3427U) | 134 [47] | 29 [47] | 181 [47] | 0.58 |
| CDC (Intel Xeon E5-2640) | 298 [47] | 58.7 [47] | 428 [47] | 0.55 |

Table 2: Networking equipment data used in the MILP model.

| Devices | Max (W) | Idle (W) | Bitrate (Gbps) | Efficiency (W/Gbps) |
|---|---|---|---|---|
| ONU Wi-Fi AP | 15 [9] | 9** | 10 [9] | 0.6 |
| OLT | 1940 [9] | 1746*** | 8600 [9] | 0.22 |
| Metro Router Port | 30 [9] | 27*** | 40 [9] | 0.08 |
| Metro Switch | 470 [9] | 423*** | 600 [9] | 0.08 |
| IP/WDM Node | 878 [9] | 790*** | 40/λ* | 0.14 |

*40Gbps / wavelength || ** is 60% of max power ||***is 90% of max power.

## III. RESULTS & DISCUSSION

We used the parameters in **Error! Reference source not found.** and **Error! Reference source not found.**, for the processing servers' and networking equipment, respectively. Where possible, device parameters have been obtained using equipment datasheets, however, we have also made simple but realistic assumptions. For example, high-capacity networking equipment located in the aggregation point of the access network, metro and core network are used by many applications and services, hence the idle power is set to 90% of the maximum power consumption. It is important to note that, we assume only a portion of the idle power consumption is associated with our application. We assume this to be 3% of the equipment's idle power consumption [9]. For low-capacity network nodes such as the ONU, we assume the idle power to be 60% of the device's maximum power consumption (hence, full idle power is consumed as these devices are deployed for our application only). In this work, we have also assumed that the centralized data centre is a single hop from the aggregation core router (aggregating from metro) and based on the topology of the NSFNET, the average distance between the core nodes is 509 km [45]. We assume that in total, there are 20 IoT devices, randomly distributed among three 10 zones. Each zone comprises of 2 IoT end-devices and these are connected via Wi-Fi links to the corresponding ONU Access Point (AP). We consider 3 OLTs in total and each OLT aggregates traffic from two ONUs. As for the workloads, we assume that each VSR has an input VM that must be mapped to the corresponding IoT device acting as the source node.

The number of VMs and the workload per hidden layer VM are randomly distributed between 2 – 4 VMs and between 2 – 13.5 GFLOPS, respectively. The input VMs workload is randomly distributed between 0.01–1 GFLOPS. As for the traffic demand, since the idle power is significantly higher than the proportional power for the networking equipment, the volume of the data rates does not influence the outcome of the optimization. We consider a PUE of 1.1 – 1.25 in AF and MF nodes, and 1.1 in the centralized CDC node. At nodes where both network and processing equipment are collocated, we assume the same PUE, however since core nodes are not collocated with processing servers in this work, we consider a PUE of 1.5 for core nodes and 1 in the remaining nodes as there is no need for cooling [9]. The adopted power profile consists of a proportional and an idle part. The proportional part increases with the volume of workload, whilst the idle part is consumed as soon as the device is activated. In the current optimization model, it is assumed that any unused equipment is switched off completely. We also assume that at each AF and MF node, a maximum of 6 and 10 servers are available, respectively. While we assume the centralized data centre nodes have unlimited number of servers. Finally, the MILP model is solved using IBM's commercial solver CPLEX over the University of Leeds high performance computing facilities (ARC3) using 24 cores with 126 GB of RAM [46].

In the following subsections, we study the CFN approach in two settings; A) when the centralized CDC is available in addition to the fog layers. and B) when the CFN is without the centralized CDC. Furthermore, for each of the aforementioned settings, we study the impact of the input/source node distribution on the performance of the CFN approach compared to the centralized CDC.

## A. CFN Approach with CDC

We evaluate the performance of the CFN approach in two scenarios; (i) when the input data of VSRs originate from a single IoT source node, i.e. one IoT device is feeding its readings into the input layer of the DNN requests (VSRs), and (ii) when the input data originates from multiple IoT devices. In this work, scenario (i) considers one input device in a single IoT zone, whilst scenario (ii) considers 10 input devices in total (1 per zone). Figure 2 shows the total power consumption versus the number of VSRs in both scenarios, compared to their corresponding CFN and baseline approaches. In the baseline approach, it is assumed that fog computing is not available, hence all VSRs are processed inside the centralized CDCs. Whereas in the CFN approach, we allow the MILP model to optimally allocate the VSRs into multiple processing layers provided by the fog. It is shown that, when compared to the baseline, during low number of VSRs (1 VSR – 9 VSRs) and for 1 IoT input node, the CFN solution achieves up to 43% power consumption reduction. This is because VSRs can be processed on local low-power IoT devices, as a result making significant savings in both processing and networking resources. On average, with our CFN solution, savings were 33% and in the worst-case scenario this dropped to 27%, which is still significant. It can be observed in Figure 5(a) that the IoT layer plays a significant role in the processing of the VSRs and that the AF and MF are never utilized. Instead, the IoT and the CDC are used in combination to host the VSRs. The majority of the workloads are allocated to the IoT layer when capacity constraints permit it (1 VSR – 10 VSRs), however, when this is no longer the case, the same is true for CDCs. Interestingly, IoT utilization increases again (from 28 VSRs – 30 VSRs). This happens to avoid activating an additional server at the CDC due to its high idle power consumption.

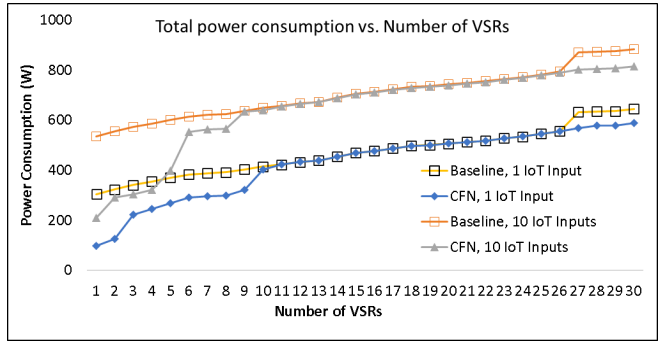

Figure 2 Total power consumption with CDC.

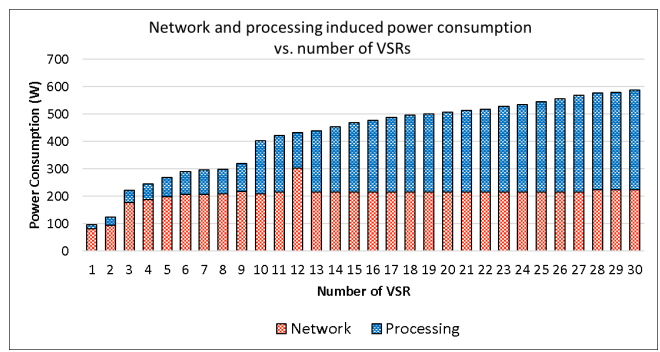

Figure 3 Breakdown of CFN's network and processing power consumption, given a total of 1 IoT input/source node.

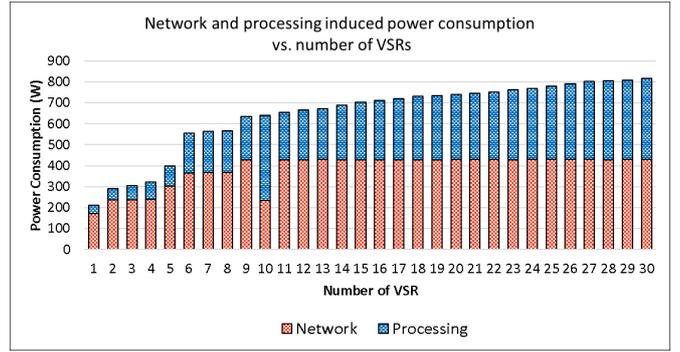

Figure 4 Breakdown of CFN's network and processing power consumption, given 10 IoT input/source node with 1 per zone.

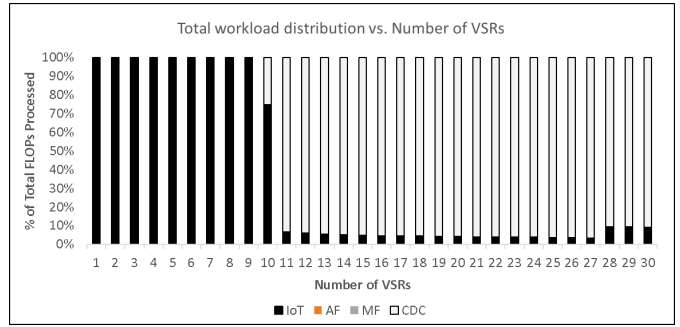

(a)

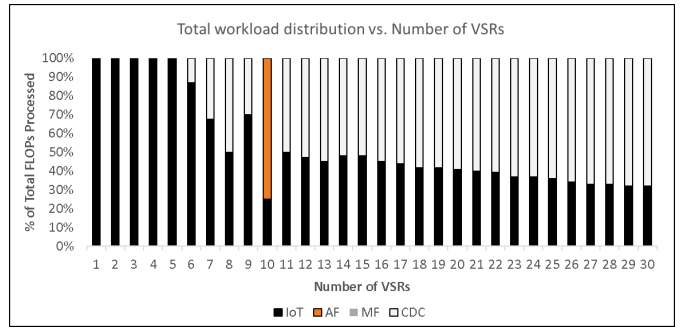

(b)

Figure 5 Workload distribution of: (a) CFN approach, given a total of 1 input, and (b) CFN approach, given 10 inputs.

In the second scenario, we adjust the input distribution to study its impact on the performance of the CFN solution compared with the corresponding baselines. As shown in Figure 2, the trends remain almost identical to the first scenario such that the IoT and CDC layers are predominantly the optimal choice (albeit limited utilization of AF layer at 10 VSRs only) of processing and the AF/MF have insignificant roles to play. This is interesting because regardless of where the VSRs are hosted, all of the OLT devices that connect the different zones will be activated as we have 1 input per zone. So, one would expect AF/MF servers that are closer to the IoT input nodes to be utilized more and accessing CDC over metro and core to be minimized. This is understandable since the objective function is purely based around power consumption and not other metrics such as latency, the model will always minimize the former regardless of the distance/ number of hops. The savings in this scenario compared to the baseline is up to 60% and at least 22%, which yields on average 10% savings. This low average saving can be attributed to the fact that input/source nodes are geographically distributed in the network, hence more networking power is consumed to establish the communication between the different VMs (or DNN layers) compared to the first scenario where a total of 1 input node existed in the whole network. We can conclude from the obtained results that, when an energy efficient CDC is available, deploying small-large fog servers in the network is not an optimal choice, unless the processing efficiency of these servers is significantly improved and PUE values are minimized.

*B. CFN Approach without CDC*

In this subsection, we aim to corroborate our claims in the previous subsection that the IoT – CDC combination is indeed the optimal choice. Hence, we run both CFN scenarios again for the case of 1 input node and 10 input nodes, however this time without the availability of the CDC node, which means VSRs can only be allocated between IoTs and AFs/MF. Figure 6 shows that, in the case where only a single input node exists in the whole network, opting for the CFN without CDC approach would yield a 4.5% increase in power consumption on average, compared to the scenario where CDC is available. As expected, in the second scenario where 10 input nodes exist, average savings moved up to 0.36% because of the instance where the AF is utilized to process about 61% of the total workload at 10 VSRs as per Figure 5(b). This can be attributed to the difference between the networking overhead of the CDC and the AF.

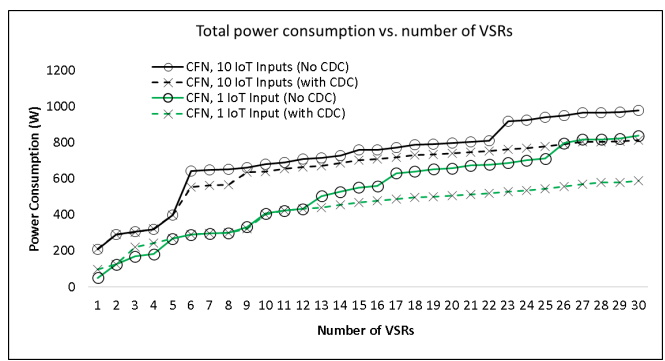

Figure 6 Total power consumption with/out CDC.

## IV. CONCLUSIONS AND FUTURE WORK

This paper introduced a flexible optimization framework which allows for DNN models to be abstracted as a virtual embedding problem in a cloud fog network architecture comprising of heterogenous layers of processing. We studied the distribution of IoT input nodes in two scenarios: 1) where a single input node existed and 2) where 10 input nodes existed. The aim of this evaluation was to substantiate the savings of the CFN approach compared with the legacy CDC processing. The results showed substantial savings, however only when workloads were allocated between IoT devices and CDC. In this work, we assumed an energy efficient CDC. It would be of interest to re-evaluate the model with an energy inefficient CDC deployed at the core network. Motivated by the obtained results, interesting research directions can include developing heuristic algorithms with improved computational complexity, considering realistic core network topologies such as the NSFNET, AT&T network and BT network, constraining IoT devices to limited power budgets and last but not least looking at the impact of renewable sources of energy at different fog sites.

ACKNOWLEDGMENT

The authors would like to acknowledge funding from the Engineering and Physical Sciences Research Council (EPSRC), INTERNET (EP/H040536/1), STAR (EP/K016873/1) and TOWS (EP/S016570/1) projects. All data are provided in full in the results section of this paper. The first author would like to thank the University of Tabuk for funding his PhD scholarship.

REFERENCES

[1] M. M. Alenazi, B. A. Yosuf, T. El-Gorashi and J. M. H. Elmirghani, "Energy Efficient Neural Network Embedding in IoT over Passive Optical Networks," *2020 22nd International Conference on Transparent Optical Networks (ICTON)*, Bari, Italy, 2020, pp. 1-6, doi: 10.1109/ICTON51198.2020.9203403.

[2] Lionel Sujay Vailshery, "Internet of Things (IoT) and non-IoT active device connections worldwide from 2010 to 2025," (Cited on May 10, 2021) [Online]. Available: https://www.statista.com/statistics/1101442/iot-number-of-connected-devices-worldwide/

[3] D. Li, X. Chen, M. Becchi and Z. Zong, "Evaluating the Energy Efficiency of Deep Convolutional Neural Networks on CPUs and GPUs," *2016 IEEE International Conferences on Big Data and Cloud Computing (BDCloud), Social Computing and Networking (SocialCom), Sustainable Computing and Communications (SustainCom) (BDCloud-SocialCom-SustainCom)*, Atlanta, GA, USA, 2016, pp. 477-484, doi: 10.1109/BDCloud-SocialCom-SustainCom.2016.76.

[4] S. L. Jurj, F. Opritoiu, and M. Vladutiu, "Environmentally-friendly metrics for evaluating the performance of deep learning models and systems," in *Neural Information Processing*, Cham: Springer International Publishing, 2020, pp. 232–244.

[5] D. C. Nguyen, M. Ding, P. N. Pathirana, A. Seneviratne, J. Li and H. V. Poor, "Federated Learning for Internet of Things: A Comprehensive Survey," in IEEE Communications Surveys & Tutorials, doi: 10.1109/COMST.2021.3075439.

[6] R. Deng, R. Lu, C. Lai, and T. H. Luan, "Towards power consumption-delay tradeoff by workload allocation in cloud-fog computing," IEEE Int. Conf. Commun., vol. 2015-Septe, pp. 3909–3914, 2015, doi:10.1109/ICC.2015.7248934.

[7] Y. Sahni, J. Cao, S. Zhang, and L. Yang, "Edge Mesh: A New Paradigm to Enable Distributed Intelligence in Internet of Things," *IEEE Access*, vol. 5, pp. 16441–16458, 2017, doi: 10.1109/ACCESS.2017.2739804.

[8] R. Mahmud, R. Kotagiri, and R. Buyya, "Fog Computing: A Taxonomy, Survey and Future Directions," pp. 1–28, 2016, doi: 10.1007/978-981-10-5861-5_5.

[9] B. A. Yosuf, S. H. Mohamed, M. M. Alenazi, T.E.H. El-Gorashi and J. M. H. Elmirghani, "Energy-Efficient AI over a Virtualized Cloud Fog Network," *2021 The 2$^{nd}$ International workshop on energy-efficent learning at the edge, WEEE 2021*.

[10] B. A. Yosuf, M. Musa, T. El-Gorashi and J. Elmirghani, "Energy Efficient Distributed Processing for IoT," in *IEEE Access*, vol. 8, pp. 161080-161108, 2020, doi: 10.1109/ACCESS.2020.3020744.

[11] Z. T. Al-Azez, A. Q. Lawey, T. E. H. El-Gorashi, and J. M. H. Elmirghani, "Energy Efficient IoT Virtualization Framework With Peer to Peer Networking and Processing," *IEEE Access*, vol. 7, pp. 50697–50709, 2019, doi: 10.1109/ACCESS.2019.2911117.

[12] H. Q. Al-Shammari, A. Q. Lawey, T. E. H. El-Gorashi, and J. M. H. Elmirghani, "Service Embedding in IoT Networks," *IEEE Access*, vol. 8, pp. 2948–2962, 2020, doi: 10.1109/ACCESS.2019.2962271.

[13] H. Q. Al-Shammari, A. Q. Lawey, T. E. H. El-Gorashi, and J. M. H. Elmirghani, "Resilient Service Embedding in IoT Networks," *IEEE Access*, vol. 8, pp. 123571–123584, 2020, doi: 10.1109/ACCESS.2020.3005936.

[14] H. A. Alharbi, T. E. H. Elgorashi, and J. M. H. Elmirghani, "Energy efficient virtual machines placement over cloud-fog network architecture," *IEEE Access*, vol. 8, pp. 94697–94718, 2020, doi: 10.1109/ACCESS.2020.2995393.

[15] A. N. Al-Quzweeni, A. Q. Lawey, T. E. H. Elgorashi, and J. M. H. Elmirghani, "Optimized Energy Aware 5G Network Function Virtualization," *IEEE Access*, vol. 7, pp. 44939–44958, 2019, doi: 10.1109/ACCESS.2019.2907798.

[16] L. Nonde, T. E. H. El-gorashi, and J. M. H. Elmirghani, "Energy Efficient Virtual Network Embedding for Cloud Networks," *Energy Effic. Virtual Netw. Embed. Cloud Networks*, vol. 33, no. 9, pp. 1828–1849, 2015, doi: 10.1109_JLT.2014.2380777.

[17] X. Dong, T. El-Gorashi, and J. M. H. Elmirghani, "Green IP over WDM networks with data centers," *J. Light. Technol.*, vol. 29, no. 12, pp. 1861–1880, 2011, doi: 10.1109/JLT.2011.2148093.

[18] X. Dong, T. El-Gorashi, and J. M. H. Elmirghani, "IP over WDM networks employing renewable energy sources," *J. Light. Technol.*, vol. 29, no. 1, pp. 3–14, 2011, doi: 10.1109/JLT.2010.2086434.

[19] N. I. Osman, T. El-Gorashi, L. Krug, and J. M. H. Elmirghani, "Energy-efficient future high-definition TV," *J. Light. Technol.*, vol. 32, no. 13, pp. 2364–2381, Jul. 2014, doi: 10.1109/JLT.2014.2324634.

[20] A. Q. Lawey, T. E. H. El-Gorashi, and J. M. H. Elmirghani, "BitTorrent content distribution in optical networks," *J. Light. Technol.*, vol. 32, no. 21, pp. 3607–3623, Nov. 2014, doi: 10.1109/JLT.2014.2351074.

[21] A. Q. Lawey, T. E. H. El-Gorashi, and J. M. H. Elmirghani, "Distributed energy efficient clouds over core networks," *J. Light. Technol.*, vol. 32, no. 7, pp. 1261–1281, 2014, doi: 10.1109/JLT.2014.2301450.

[22] H. A. Alharbi, T. E. H. Elgorashi, and J. M. H. Elmirghani, "Impact of the Net Neutrality Repeal on Communication Networks," *IEEE Access*, vol. 8, pp. 59787–59800, 2020, doi: 10.1109/ACCESS.2020.2983314.

[23] S. H. Mohamed, M. B. A. Halim, T. E. H. Elgorashi, and J. M. H. Elmirghani, "Fog-Assisted Caching Employing Solar Renewable Energy and Energy Storage Devices for Video on Demand Services," *IEEE Access*, vol. 8, pp. 115754–115766, 2020, doi: 10.1109/ACCESS.2020.3004314.

[24] A. M. Al-Salim, A. Q. Lawey, T. E. H. El-Gorashi, and J. M. H. Elmirghani, "Energy Efficient Big Data Networks: Impact of Volume and Variety," *IEEE Trans. Netw. Serv. Manag.*, vol. 15, no. 1, pp. 458–474, Mar. 2018, doi: 10.1109/TNSM.2017.2787624.

[25] A. M. Al-Salim, T. E. H. El-Gorashi, A. Q. Lawey, and J. M. H. Elmirghani, "Greening big data networks: Velocity impact," *IET Optoelectron.*, vol. 12, no. 3, pp. 126–135, Jun. 2018, doi: 10.1049/iet-opt.2016.0165.

[26] M. Hadi, A. Lawey, T. El-Gorashi, and J. Elmirghani, "Using Machine Learning and Big Data Analytics to Prioritize Outpatients in HetNets," in *IEEE INFOCOM 2019 - IEEE Conference on Computer Communications Workshops (INFOCOM WKSHPS)*, 2019, pp. 726–731, doi: 10.1109/INFCOMW.2019.8845225.

[27] M. S. Hadi, A. Q. Lawey, T. E. H. El-Gorashi, and J. M. H. Elmirghani, "Patient-Centric Cellular Networks Optimization Using Big Data Analytics," *IEEE Access*, vol. 7, pp. 49279–49296, 2019, doi: 10.1109/ACCESS.2019.2910224.

[28] M. S. Hadi, A. Q. Lawey, T. E. H. El-Gorashi, and J. M. H. Elmirghani, "Patient-Centric HetNets Powered by Machine Learning and Big Data Analytics for 6G Networks," *IEEE Access*, vol. 8, pp. 85639–85655, 2020, doi: 10.1109/ACCESS.2020.2992555.

[29] I. S. B. M. Isa, T. E. H. El-Gorashi, M. O. I. Musa, and J. M. H. Elmirghani, "Energy Efficient Fog-Based Healthcare Monitoring Infrastructure," *IEEE Access*, vol. 8, pp. 197828–197852, 2020, doi: 10.1109/ACCESS.2020.3033555.

[30] B. G. Bathula, M. Alresheedi, and J. M. H. Elmirghani, "Energy Efficient Architectures for Optical Networks," in *Proceedings IEEE London Communications Symposium, London,* 2009, pp. 5–8, Accessed: Jan. 08, 2020. [Online]. Available: http://www.ee.ucl.ac.uk/lcs/previous/LCS2009/LCS/lcs09_33.pdf.

[31] B. G. Bathula and J. M. H. Elmirghani, "Energy efficient Optical Burst Switched (OBS) networks," 2009, doi: 10.1109/GLOCOMW.2009.5360734.

[32] X. Dong, T. E. H. El-Gorashi, and J. M. H. Elmirghani, "On the energy efficiency of physical topology design for IP over WDM networks," *J. Light. Technol.*, vol. 30, no. 12, pp. 1931–1942, 2012, doi: 10.1109/JLT.2012.2186557.

[33] X. Dong, A. Lawey, T. E. H. El-Gorashi, and J. M. H. Elmirghani, "Energy-efficient core networks," 2012, doi: 10.1109/ONDM.2012.6210196.

[34] T. E. H. El-Gorashi, X. Dong, and J. M. H. Elmirghani, "Green optical orthogonal frequency-division multiplexing networks," *IET Optoelectron.*, vol. 8, no. 3, pp. 137–148, 2014, doi: 10.1049/iet-opt.2013.0046.

[35] J. M. H. Elmirghani et al., "GreenTouch GreenMeter Core Network Energy Efficiency Improvement Measures and Optimization [Invited]," *IEEE/OSA J. Opt. Commun. Netw.*, vol. 10, no. 2, 2018, doi: 10.1364/JOCN.10.00A250.

[36] M. O. I. Musa, T. E. H. El-Gorashi, and J. M. H. Elmirghani, "Bounds on GreenTouch GreenMeter Network Energy Efficiency," *J. Light. Technol.*, vol. 36, no. 23, pp. 5395–5405, Dec. 2018, doi: 10.1109/JLT.2018.2871602.

[37] M. Musa, T. Elgorashi, and J. Elmirghani, "Bounds for energy-efficient survivable IP over WDMnetworks with network coding," *J. Opt. Commun. Netw.*, vol. 10, no. 5, pp. 471–481, May 2018, doi: 10.1364/JOCN.10.000471.

[38] M. Musa, T. Elgorashi, and J. Elmirghani, "Energy efficient survivable IP-Over-WDM networks with network coding," *J. Opt. Commun. Netw.*, vol. 9, no. 3, pp. 207–217, Mar. 2017, doi: 10.1364/JOCN.9.000207.

[39] H. M. M. Ali, T. E. H. El-Gorashi, A. Q. Lawey, and J. M. H. Elmirghani, "Future Energy Efficient Data Centers with Disaggregated Servers," *J. Light. Technol.*, vol. 35, no. 24, pp. 5361–5380, Dec. 2017, doi: 10.1109/JLT.2017.2767574.

[40] C. Gray, R. Ayre, K. Hinton, and R. S. Tucker, "Power consumption of IoT access network technologies," 2015 IEEE Int. Conf. Commun. Work., pp. 2818–2823, 2015, doi: 10.1109/ICCW.2015.7247606.

[41] C. Mouradian, D. Naboulsi, S. Yangui, R. H. Glitho, M. J. Morrow, and P. A. Polakos, "A Comprehensive Survey on Fog Computing: State-of-the-Art and Research Challenges," *IEEE Commun. Surv. Tutorials*, vol. 20, no. 1, pp. 416–464, 2018, doi: 10.1109/COMST.2017.2771153.